\documentclass[prd,amssymb,amsmath,amsfonts,aps,altaffilletter]{revtex4}
\usepackage{subfigure}
\usepackage[pdftex]{graphicx}

\begin{document}

\begin{abstract}
We summarize the sensitivity achieved by the LIGO and Virgo gravitational wave detectors for compact binary coalescence (CBC) searches during LIGO's fifth science run and Virgo's first science run.  We present noise spectral density curves for each of the four detectors that operated during these science runs which are representative of the typical performance achieved by the detectors for CBC searches.  These spectra are intended for release to the public as a summary of detector performance for CBC searches during these science runs.
\end{abstract}

\title{Sensitivity to Gravitational Waves from Compact Binary Coalescences Achieved during LIGO's Fifth and Virgo's First Science Run\\
$ $\\
LIGO T0900499 v19\\
VIR 0171A 10}
\author{J.~Abadie$^{29}$, 
B.~P.~Abbott$^{29}$, 
R.~Abbott$^{29}$, 
M,~Abernathy$^{66}$, 
T.~Accadia$^{27}$, 
F.~Acernese$^{19ac}$, 
C.~Adams$^{31}$, 
R.~Adhikari$^{29}$, 
P.~Ajith$^{29}$, 
B.~Allen$^{2,78}$
% ...
\\
(The LIGO Scientific Collaboration and the Virgo Collaboration)\\
and K.~Belczynski$^{80,81}$}
% ...
\address{$^{2}$Albert-Einstein-Institut, Max-Planck-Institut f\"ur Gravitationsphysik, D-30167 Hannover, Germany}
% ...
\address{$^{19}$INFN, sezione di Napoli $^a$; Universit\`a di Napoli 'Federico II'$^b$ Complesso Universitario di Monte S.Angelo, I-80126 Napoli; Universit\`a di Salerno, Fisciano, I-84084 Salerno$^c$, Italy}
% ...
\address{$^{27}$Laboratoire d'Annecy-le-Vieux de Physique des Particules (LAPP),  IN2P3/CNRS, Universit\'e de Savoie, F-74941 Annecy-le-Vieux, France}
% ...
\address{$^{29}$LIGO - California Institute of Technology, Pasadena, CA  91125, USA }
% ...
\address{$^{31}$LIGO - Livingston Observatory, Livingston, LA  70754, USA }
% ...
\address{$^{66}$University of Glasgow, Glasgow, G12 8QQ, United Kingdom }
% ...
\address{$^{78}$University of Wisconsin--Milwaukee, Milwaukee, WI  53201, USA }
% ...
\address{$^{80}$Los Alamos National Laboratory, CCS-2/ISR-1 Group, Los Alamos, NM, USA }
\address{$^{81}$Astronomical Observatory, University of Warsaw, Al.~Ujazdowskie 4, 00-478 Warsaw, Poland }
% ...

\maketitle

\section{Introduction}
In this note, we summarize the sensitivity to gravitational waves from compact binary coalescences (CBCs) achieved during LIGO's fifth science run (S5) and Virgo's first science run (VSR1) \cite{Abbott:2009,Acernese:2008}.  A complete and search-independent characterization of the sensitivity of a gravitational wave detector over short timescales is given by the spectral density.  Over long timescales, the noise in LIGO and Virgo detectors is non-stationary and a spectral density is not an appropriate description of a detector's sensitivity.  Our goal here is to characterize the overall performance of each detector for CBC searches in all of S5 and VSR1 using the familiar notion of spectral density.  However, since the concept of spectral density is not appropriate for LIGO and Virgo detectors on such a long timescale, we choose a single ``representative'' time for each detector and compute the spectral density in each detector around this time.  We take the resulting spectral density as representative of the typical sensitivity to gravitational waves from CBCs achieved by the detectors in S5/VSR1.

We choose the representative spectral density curves to correspond to times when each detector operated near the mode of its inspiral horizon distance distribution.  The inspiral horizon distance is a quantity derived from the spectral density that summarizes the sensitivity of a detector to gravitational waves from CBCs at a given time.  In this article, we gather the inspiral horizon distance data generated during S5/VSR1 CBC analysis and use the results to identify spectral density curves that are representative of detector performance for CBC searches during these science runs.

The plots and data presented here are intended to be released to the public as a summary of detector performance for CBC searches during S5 and VSR1.  These results use exactly the same science segments and analysis code \cite{TmpltBank} that was used in the CBC searches in S5 and VSR1.  The results presented here supersede previous studies \cite{SiggDutyCycle} done on the inspiral horizon distance in that we use LIGO's version 3 calibration \cite{Goetz:2009}, the same calibration version used in the S5 CBC search.  Using version 4 calibration \cite{S5V4Calibration} would change these results by about 5 to 10\%, but not in a simple way.  Since version 4 $h(t)$ has never been made for S5 and version 3 calibration was used in the S5 search, we stick with version 3 calibration here.  However, we do include first order corrections for version 4 DC calibration, which affects the results through an overall multiplicative scaling factor.  All CBC searches applied these DC calibration corrections, so the results presented here compare directly with the CBC search sensitivities reported in our papers.  For Virgo, we use version 2 calibration \cite{VirgoCalibration}, the same calibration used in the VSR1 CBC search.

In the next section, we define the inspiral horizon distance and present a summary of the inspiral horizon distance data from S5/VSR1 CBC analyses.  In section 3, we explain how we used the inspiral horizon distance data to compute a spectral density that is representative of each detector's sensitivity for CBC searches in  S5/VSR1.

\section{Inspiral Horizon Distance}

The (power) spectral density $S_n(f)$ for a stationary random process $n(t)$ is defined implicitly by the relation
\begin{eqnarray}
\label{psd}\frac{1}{2} S_n(f)\delta(f-f') =  \langle \tilde{n}(f)\tilde{n}^*(f') \rangle ,
\end{eqnarray}
where $\tilde{n}(f)$ is the Fourier transform of the random process.  The spectral density is a measure of the mean square noise fluctuations at a given frequency.   In LIGO and Virgo applications, we treat the strain noise in a detector as a stationary random process.  If the noise in the detector were truly stationary, then the noise spectral density would completely characterize the sensitivity of the detector as a function of frequency.

As mentioned above, the noise in the LIGO and Virgo detectors is not stationary.  However, by measuring the spectral density over a short enough timescale, we are able to approximate the noise as stationary.  The chosen timescale must also be long enough that we can form an accurate estimate of the spectral density.  In the S5/VSR1 CBC searches, the spectral density was computed on 2048-second blocks of contiguous data \cite{FindChirp}.  We account for long timescale non-stationarities by using a different spectral density for every 2048 seconds.

In assessing the overall performance of a detector for CBC searches, we use the inspiral horizon distance data from S5 and VSR1 to identify the ``typical'' sensitivity of the interferometers.  The inspiral horizon distance of a detector is the distance at which an optimally oriented and optimally located equal-mass compact binary inspiral would give an average signal to noise ratio (SNR) of $\rho=8$ in the interferometer. If $\tilde{h}(f)$ represents the Fourier transform of the expected signal, then the average SNR this signal would attain in a detector with spectral density $S_n(f)$ is given by
\begin{eqnarray}
\label{general_ir} \langle \rho \rangle= \sqrt{4 \int_0^\infty \frac{|\tilde{h}(f)|^2}{S_n(f)} df}.
\end{eqnarray}
We find the inspiral horizon distance by setting $\langle \rho \rangle = 8$ and solving for the distance $D$ to the inspiral event which parametrizes the waveform $\tilde{h}(f)$.  Thus, the inspiral horizon distance combines the spectral density curve with the expected inspiral waveform to produce a single quantity that summarizes the sensitivity of the detector at a given time.

Practical considerations require modifications to the limits of the integral.  In the CBC search code, we compute the signal to noise ratio by
\begin{eqnarray}
\label{range} \langle \rho \rangle= \sqrt{4 \int_{f_{low}}^{f_{high}} \frac{|\tilde{h}(f)|^2}{S_n(f)} df}.
\end{eqnarray}
The lower limit is determined by our ability to characterize the noise at low frequencies.  In the S5 CBC search, we took $f_{low}=40$Hz as the low frequency cut-off in computing the inspiral horizon distance.  For Virgo in VSR1, the low frequency cut-off was $f_{low}=60$Hz.  The upper limit of the integral is the innermost stable circular orbit (ISCO) frequency,
\begin{eqnarray}
f_{isco} = \frac{c^3}{6\sqrt{6}\pi G M},
\end{eqnarray}
where $M$ is the total mass of the binary system.  For binary neutron star systems, $f_{isco}= 1570$Hz.  However, the inspiral horizon distances reported here are culled from the S5 high mass search (except for Virgo), which down-sampled the $h(t)$ data to 2048 Hz, so that the integral was cut off at $f_{Ny}$ = 1024 Hz. However, this is a $<1\%$ effect, even in computing the inspiral horizon distance for low mass systems, because most of the SNR is accumulated in the ``bucket'' of the noise curve; we neglect it here.

The inspiral waveform for CBCs is accurately given in the frequency domain by the stationary phase approximation.  For an optimally oriented and optimally located equal mass binary, the signal that appears at the interferometer (in this approximation) is given by
\begin{eqnarray}
\label{spa}
\tilde{h}(f) =  \frac{1}{D}\left(\frac{5\pi }{24c^3}\right)^{1/2}(G\mathcal{M})^{5/6}(\pi f)^{-7/6} e^{i\Psi(f;M)},
\end{eqnarray}
where $\mathcal{M}$ is the chirp mass of the binary, $D$ is the distance to the binary and $\Psi$ is a real function of $f$, parametrized by the total mass $M$.  Setting $\langle \rho \rangle = 8$ and inserting this waveform into eqn. \ref{range}, we find that the inspiral horizon distance is given by
\begin{eqnarray}
\label{range0} D = \frac{1}{8}\left(\frac{5\pi }{24c^3}\right)^{1/2}(G\mathcal{M})^{5/6}\pi^{-7/6} \sqrt{4 \int_{f_{low}}^{f_{high}} \frac{f^{-7/3}}{S_n(f)}df },
\end{eqnarray}
where $D$ is expressed in Mpc.  The inspiral horizon distance is defined for optimally located and oriented sources. For a uniform distribution of source sky locations and orientations, we divide the inspiral horizon distance by 2.26 to obtain the SenseMon range \cite{Finn:1993} reported as a figure of merit in the LIGO and Virgo control rooms.

In practice, it is convenient to measure distances in Mpc and  mass in $M_\odot$. It is useful therefore to specialize eqn. \ref{range0} to this unit system.  Further, since we measure the strain $h(t)$ at discrete time intervals $\Delta t = 1/f_s$, the spectral density is only known with a frequency resolution of $\Delta f = f_{s}/N$, where $N$ is the number of data points used to measure $S_n(f)$.  By putting $f=k\Delta t$ into eqn. \ref{range0} and grouping terms by units, we arrive at the expression
\begin{eqnarray}
 D \approx \frac{1}{8} \mathcal{T} \sqrt{\frac{4}{N} \sum_{k_{low}}^{k=k_{high}} \frac{(k/N)^{-7/3}}{S_n(k)} } \mathrm{ Mpc },
\end{eqnarray}
where
\begin{eqnarray}
\mathcal{T} = \left( \frac{5}{24\pi^{4/3}} \right)^{1/2}
              \left( \frac{ \mu }{ M_\odot } \right)^{1/2}
              \left( \frac{ M }{ M_\odot } \right)^{1/3}
              \left( \frac{ G M_{\odot}/c^2 }{\mathrm{1 Mpc}} \right)
              \left( \frac{ G M_\odot/c^3 }{\Delta t } \right)^{-1/6},
\end{eqnarray}
for the inspiral horizon distance in Mpc. Since it is convenient to work with the binary system's component masses, we have also replaced the chirp mass $\mathcal{M}$ with the reduced mass $\mu$ and the total mass $M$, where $\mathcal{M} = \mu^{3/5}M^{2/5}$.  Written this way, the inspiral horizon distance in Mpc is easily computed from the binary component masses in $M_\odot$.

\begin{figure}
\includegraphics[scale=0.65]{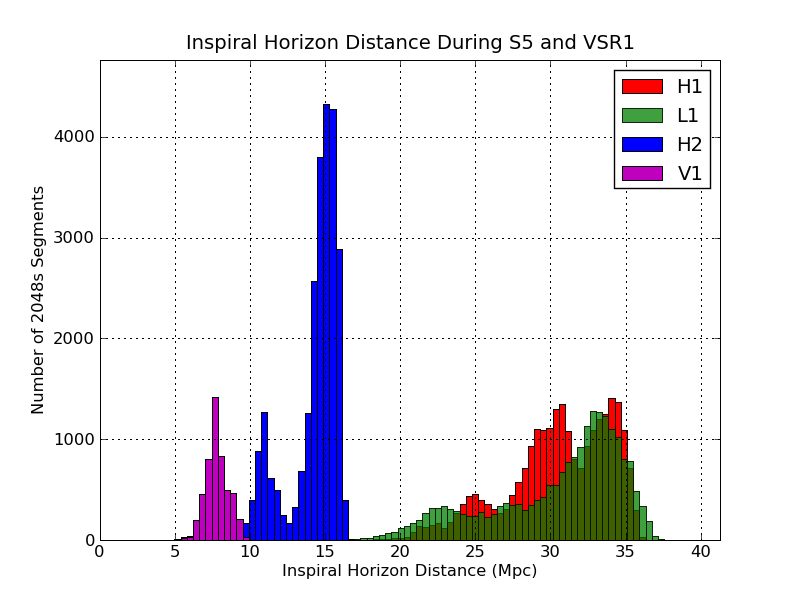}
\caption{Distribution of inspiral horizon distance for the four gravitational wave detectors H1, L1, H2 and V1 for all of S5 and VSR1.  This histogram includes each 2048-second analyzed segment from S5 and VSR1.  The distributions shown here correspond to the 1.4 -1.4 solar mass inspiral horizon distance for the LIGO detectors.  For the Virgo detector, we have plotted the 1.0-1.0 solar mass inspiral horizon distance distribution, scaled by $(2.8/2)^{5/6}$ to adjust for the lower mass. }
\label{rangehist}
\end{figure}

\begin{figure}
\includegraphics[scale=0.65]{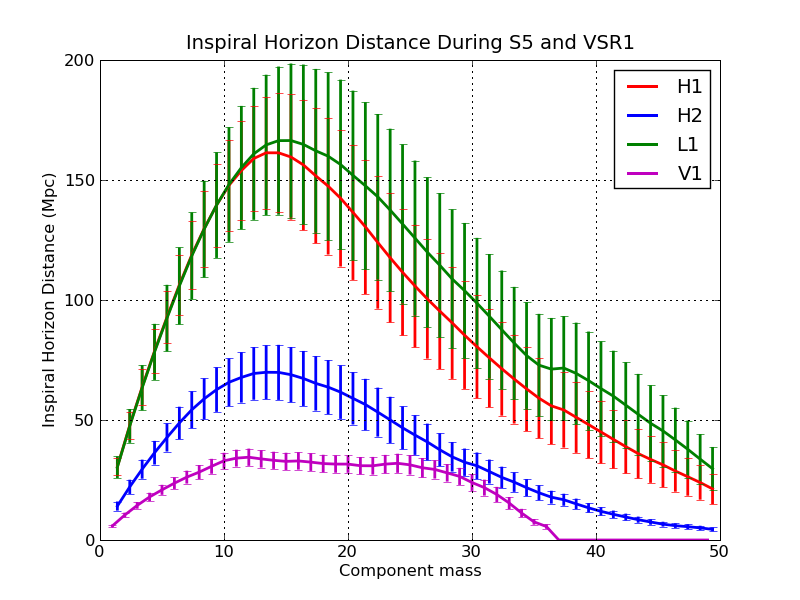}
\caption{Mean inspiral horizon distance as a function of mass for the four gravitational wave detectors H1, L1, H2 and V1 during all of S5 and VSR1.  The error bars on the curves extend from one standard deviation below to one standard deviation above the mean.}
\label{rangevmass}
\end{figure}

\begin{figure}
\includegraphics[scale=0.65]{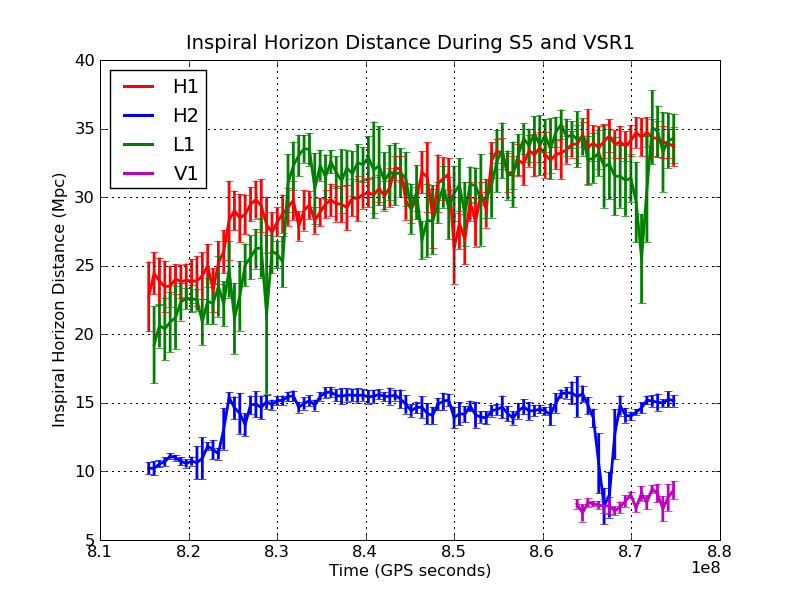}
\caption{Inspiral horizon distance as a function of time during S5 and VSR1.  The average inspiral horizon distances for each week in S5 and VSR1.  As an indication of the weekly variations, we have included error bars corresponding to the standard deviation of the inspiral horizon distance during each week.}
\label{h1curves}
\end{figure}

We have collected the inspiral horizon distance data from the original output of the S5 high-mass and S5/VSR1 low-mass CBC searches to be published.  We have collected the data, rather than computing the inspiral horizon distance directly, in order to ensure that we analyze the exact same science segments and use the exact same analysis code as used in the LIGO/Virgo CBC searches.

In fig. \ref{rangehist}, we histogram the inspiral horizon distance for the four gravitational wave detectors H1, L1, H2 and V1 for all of S5 and VSR1. The plot depicts the variability in sensitivity to gravitational waves from CBCs throughout the science runs.  The bimodal behavior seen in the LIGO detector distributions is due to a significant commissioning break in S5, which greatly improved the sensitivity of the LIGO detectors.

In fig. \ref{rangevmass}, we plot the mean inspiral horizon distance for each interferometer as a function of the binary component mass.  This plot reflects the mean performance of the detector over various frequency bands.  As the component mass becomes higher, the upper cutoff frequency $f_{high}=f_{isco}$ becomes smaller and smaller.  This means that the inspiral horizon distance focuses on a narrower band around the lower cutoff $f_{low}=40$Hz (or $f_{low}=60$Hz in the case of Virgo).  The fall-off of the inspiral horizon distance for high-mass binaries characterizes the performance of the detector near the low cut-off low frequency.  In particular, it should be emphasized that the sensitivity of the detectors during S5 and VSR1 to high-mass systems does not fall off as might be suggested by the graph.  The inspiral horizon distance takes into account only the inspiral stage of the CBC event, while for high-mass systems ($M> 25M_{sun}$) the merger and ringdown stages of the occur in the LIGO and Virgo sensitive band.  For these binary systems, we use EOBNR waveform templates that include the merger and ringdown stages and our sensitivity is significantly greater than depicted here.

For purely historical reasons, the S5 high-mass runs, which did not include V1, computed the inspiral horizon distance for ($n$+0.4)-($n$+0.4) solar mass binaries for integers $n\geq 1$.  The S5 low-mass runs, which included V1, instead computed the horizon for $n$-$n$ solar mass binaries.  In order to make an apple-to-apples comparison, we scaled the Virgo distribution by $(2.8/2)^{5/6}$ corresponding to the ratio of chirp masses for the LIGO and Virgo data.  This scaling ignores the fact that $f_{isco}$ is different for the two mass pairs, but this is negligible since the template is buried in the noise at that high of a frequency.

\section{Representative Noise Spectral Density}

In this section, we present spectral densities which we suggest are representative of the sensitivity achieved for S5/VSR1 CBC searches.  The chosen representative curve corresponds to a time when the detector operated near the mode of its inspiral horizon distance distribution.  We also illustrate the variability in detector sensitivity by giving spectra for H1 corresponding to times when H1 operated near the mode, early-S5 mode, mean and max of its inspiral horizon distance distribution.

The algorithm used to compute the spectral densities is described in detail in \cite{FindChirp}.  The parameters needed in order to reconstruct our results and are given in table \ref{params}.  The first column in table \ref{params} gives a list of parameter names and symbols, which are the same names and symbols used in \cite{FindChirp}.  The second and third columns gives the values of these parameters used in S5/VSR1 CBC searches.  These parameters can be used to reproduce the inspiral horizon distance data accompanying this note.  The fourth column gives the values of the parameters used to compute the representative spectral density curves shown here.  In making our choice of parameters for computing representative spectra, we sacrificed frequency resolution ($\Delta f = 1/T$) for PSD accuracy (which increases with $N_S$).

\begin{table}
\begin{center}
\caption{Parameters used in the computation of the spectral density.}
\label{params}
$ $\\
\begin{tabular}{l|c|c|c}
FINDCHIRP parameter \cite{FindChirp}  & S5 high-mass & VSR1 low-mass & representative spectra\\
\hline
sample rate ($1/\Delta t$) & 2048 Hz & 4096 Hz & 16384 Hz\\
data block duration ($T_{block}$) & 2048s & 2048s &  2048s\\
number of data segments ($N_S$) & 15 & 15 & 1023\\
data segment duration ($T$) & 256s & 256s & 4s\\
stride ($\Delta$) & 262144 & 524288 & 32768
\end{tabular}
\end{center}
\end{table}

\begin{figure}
\includegraphics[scale=0.65]{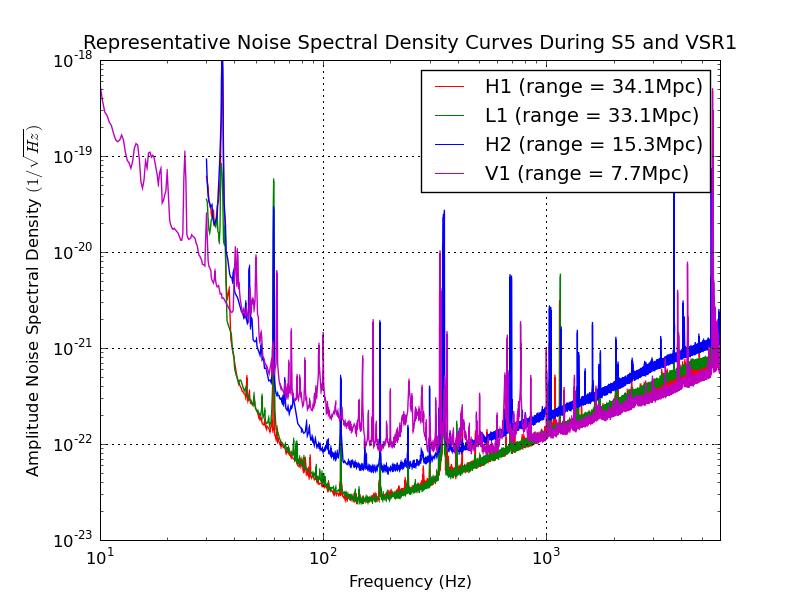}
\caption{Representative spectral density curves for LIGO and Virgo detectors during S5 and VSR1.  These spectral density curves correspond to July 21, 2007 (GPS 869040003) for H1, March 16, 2007 (GPS 858087370) for L1, September 30, 2007 (GPS 875211248) for H2 and June 13, 2007 (GPS 865748914) for V1.  These times are chosen such that the inspiral horizon distance for each detector at that time coincides with the mode of the its inspiral horizon distance distribution, as given by the midpoint of the most populated bin in fig. \ref{rangehist}.}
\label{psdcurves}
\end{figure}

In fig. \ref{psdcurves}, we give representative spectral density curves for each of the four detectors during S5 and VSR1.  These curves correspond to times when the detectors operated near the mode of their 1.4-1.4 solar mass inspiral horizon distance distribution.  The strain calibration is valid for LIGO above 30 Hz and for Virgo above 10 Hz, which give the lower limits for the curves plotted here.

In table \ref{stats}, we provide a quantitative summary of the low-mass inspiral horizon distance distributions.  All distances are given in Mpc.  The Virgo data was for 1.0-1.0 solar mass but has again been rescaled by $(2.8/2)^{5/6}$.  These statistics give varied viewpoints on the detector performance.  In the case of L1, for instance, the mode and max differ by more than 10\%.  This fact suggests that the spectral density curves corresponding times when L1 operators at its mode differ significantly from its ``best'' spectral density.

To illustrate this point, we plot in fig. \ref{h1curves} four spectra for H1 from different times in S5.  We see a significant improvement in the spectral density from the beginning of S5 to the end of S5.  In early S5, H1 operated most often with a 1.4-1.4 solar mass inspiral horizon distance near 26.7 Mpc.  In late S5, the inspiral horizon distance distribution peaked around 36.6 Mpc (early S5 roughly corresponds to the lower peaks in the horizon distributions from fig. \ref{rangehist}, while late S5 roughly corresponds to the higher peaks).  Correspondingly, we see a decrease in the spectral density from early S5 to late S5.

\begin{figure}
\includegraphics[scale=0.65]{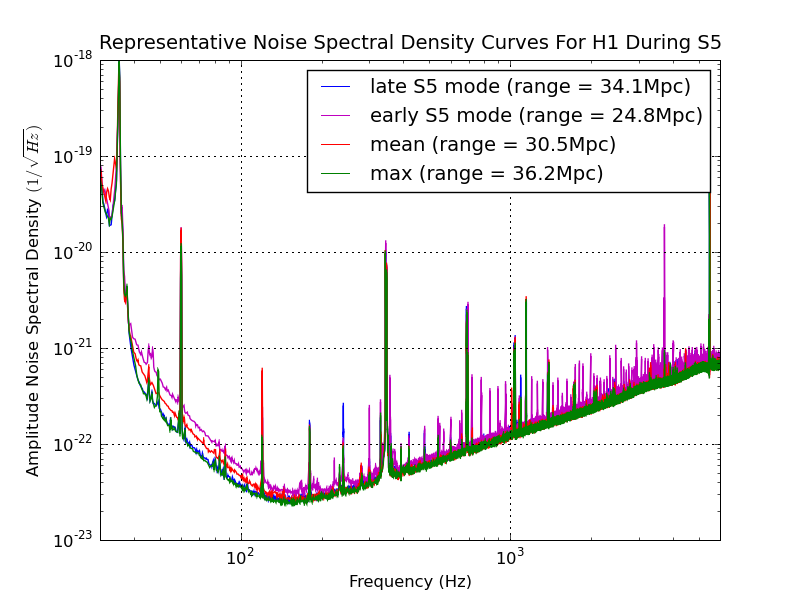}
\caption{Representative spectral density curves for H1 during S5.  These spectral density curves correspond to times when the detector operated near its late S5 mode (34.1Mpc), early S5 mode (24.8Mpc), mean (30.5Mpc) and maximum (36.2Mpc) inspiral horizon distance.  The times chosen are July 21, 2007 (GPS 869040003) for the late S5 mode, February 05, 2005 (GPS 823205705) for the early S5 mode, January 25, 2007 (GPS 853767368) for the mean and August 15, 2007 (GPS 871198828) for the maximum.}
\label{h1curves}
\end{figure}

\begin{table}
\begin{center}
\caption{Summary of Inspiral Horizon Distance Data from S5 and VSR1}
\label{stats}
$ $\\
\begin{tabular}{l|cccc}
           & H1 & L1 & H2 & V1\\
\hline
mean & 30.4 & 30.3 & 14.1 & 7.8\\
max & 36.2 & 37.5 & 16.7 & 9.8\\
mode & 34.1 & 33.1 & 15.3 & 7.7\\
std & 3.5 & 4.5 & 1.9 & 0.6
\end{tabular}
\end{center}
\end{table}

All of the LIGO data used here has been computed using version 3 calibration.  Updating these results using version 4 calibration would be nontrivial.  However, we have corrected all plots and data presented here for changes in the DC calibration from version 3 to version 4, as reported \cite{S5V4Calibration}.  This correction amounts to multiplying the spectra for each detector by an overall scaling factor.  The scaling factors for the various detectors are 1.07 for H1, 1.062 for H2 and 0.96 for L1.  The inspiral ranges are reduced by this same factor.  These corrections make H1 and H2 are somewhat less sensitive in version 4 calibration, while L1 is somewhat more sensitive.  Note that the noise spectra presented here are subject to systematic uncertainties associated with the strain calibration. These uncertainties can be up to $\pm$15\% in amplitude. For more detail, see references \cite{Goetz:2009,VirgoCalibration}.

\section{Conclusions}

We have presented spectral density curves for each of the four detectors used in S5 and VSR1.  We suggest that these noise curves are representative of the sensitivity of the LIGO and Virgo detectors for CBC searches in S5 and VSR1.  Our choice for these noise curves is based on the inspiral horizon distance, which is a measure of detector performance for binary inspirals.  We intend for these noise curves to be released to the public as a summary of detector performance during S5 and VSR1.

\section*{Acknowledgements}

The authors gratefully acknowledge the support of the United States National Science Foundation for the construction and operation of the LIGO Laboratory, the Science and Technology Facilities Council of the United Kingdom, the Max-Planck-Society, and the State of Niedersachsen/Germany for support of the construction and operation of the GEO600 detector, and the Italian Istituto Nazionale di Fisica Nucleare and the French Centre National de la Recherche Scientiﬁque for the construction and operation of the Virgo detector. The authors also gratefully acknowledge the support of the research by these agencies and by the Australian Research Council, the Council of Scientiﬁc and Industrial Research of India, the Istituto Nazionale di Fisica Nucleare of Italy, the Spanish Ministerio de Educaci´on y Ciencia, the Conselleria d’Economia Hisenda i Innovaci´o of the Govern de les Illes Balears, the Foundation for Fundamental Research on Matter supported by the Netherlands Organisation for Scientiﬁc Research, the Royal Society, the Scottish Funding Council, the Polish Ministry of Science and Higher Education, the FOCUS Programme of Foundation for Polish Science, the Scottish Universities Physics Alliance, The National Aeronautics and Space Administration, the Carnegie Trust, the Leverhulme Trust, the David and Lucile Packard Foundation, the Research Corporation, and the Alfred P. Sloan Foundation. LIGO Document No. LIGO-T0900499.

Anyone using the information in this document and associated material (S5 noise spectra, inspiral ranges, observation times) in a publication or talk must acknowledge the US National Science Foundation, the LIGO Scientific Collaboration, and the Virgo Collaboration.  Data files associated with the results and plots presented in this document can be found here: https://dcc.ligo.org/cgi-bin/private/DocDB/ShowDocument?docid=6314.  Please direct all questions to the corresponding author (sprivite@ligo.caltech.edu).  Please inform the corresponding author and the LSC and Virgo spokespeople (currently, reitze@phys.ufl.edu and francesco.fidecaro@df.unipi.it, respectively) if you intend to use this information in a publication.

\clearpage

\bibliography{S5_inspiral_range}{}

\begin{thebibliography}{1}

\bibitem{Abbott:2009}
{The LIGO Scientific Collaboration: B. Abbott, et. al.}
\newblock {LIGO: The Laser Interferometer Gravitational-Wave Observatory},
  2009.
\newblock arXiv:0711.3041.

\bibitem{Acernese:2008}
{F. Acernese, et. al.}
\newblock {\em Class. Quantum Grav.}, 25:114045, 2008.

\bibitem{TmpltBank}
{See lines 964-990 in the lalapps tmpltbank source code at\\
  http://www.lsc-group.phys.uwm.edu/cgit/lalsuite/tree/lalapps/src/inspiral/tm%
pltbank.c\\ and the function compute\_candle\_distance in lines 86-119 from\\
  http://www.lsc-group.phys.uwm.edu/cgit/lalsuite/tree/lalapps/src/inspiral/in%
spiralutils.c}.

\bibitem{SiggDutyCycle}
Daniel Sigg.
\newblock {Range and Duty Cycle}.
\newblock {LIGO T0900503-v1,
  https://dcc.ligo.org/cgi-bin/DocDB/ShowDocument?docid=6418}.

\bibitem{Goetz:2009}
{E. Goetz, et. al.}
\newblock {Accurate calibration of test mass displacement in the LIGO
  interferometers, for the proceedings of 8th Edoardo Amaldi Conference on
  Gravitational Waves}, 2009.
\newblock {arXiv:0911.0853}.

\bibitem{S5V4Calibration}
{http://touro.ligo-la.caltech.edu/$\sim$irish/Work/Calibration/S5V4Review/Over%
view.html}.

\bibitem{VirgoCalibration}
F.~Marion, B.~Mours, and L.~Rolland.
\newblock {$h(t)$ reconstruction for VSR1; Version 2 and 3}.
\newblock {VIR-0078A-08,
  https://pub3.ego-gw.it/itf/tds/index.php?callContent=2\&callCode=2092}.

\bibitem{FindChirp}
Bruce Allen, Warren~G. Anderson, Patrick~R. Brady, Duncan~A. Brown, and Jolien
  D.~E. Creighton.
\newblock {FINDCHIRP: an algorithm for detection of gravitational waves from
  inspiraling compact binaries}.
\newblock arXiv:gr-qc/0509116.

\bibitem{Finn:1993}
Lee~Samuel Finn and David~F. Chernoff.
\newblock {Observing binary inspiral in gravitational radiation: One
  interferometer}.
\newblock {\em {Phys. Rev. D}}, 47:2198--2219, 1993.

\end{thebibliography}
\bibliographystyle{unsrt}

% FIXME: Need textual reference for Acernese2008

\end{document}